\pdfoutput=1
\documentclass[sigconf]{acmart}

\def\BibTeX{{\rm B\kern-.05em{\sc i\kern-.025em b}\kern-.08emT\kern-.1667em\lower.7ex\hbox{E}\kern-.125emX}}
    
\usepackage{tabularx,ragged2e}
\usepackage{multirow}
\usepackage{subcaption}
\usepackage{balance}

\setcopyright{acmcopyright}
\copyrightyear{2023}
\acmYear{2023}
\setcopyright{acmlicensed}\acmConference[MM '23]{Proceedings of the 31st ACM International Conference on Multimedia}{October 29-November 3, 2023}{Ottawa, ON, Canada}
\acmBooktitle{Proceedings of the 31st ACM International Conference on Multimedia (MM '23), October 29-November 3, 2023, Ottawa, ON, Canada}
\acmPrice{15.00}
\acmDOI{10.1145/3581783.3613853}
\acmISBN{979-8-4007-0108-5/23/10}

\begin{document}

%

\title[SMP Challenge]{SMP Challenge: An Overview and Analysis of Social Media Prediction Challenge}

\author{Bo Wu}
\affiliation{%
  \institution{MIT-IBM Watson AI Lab}
}

\author{Peiye Liu}
\affiliation{%
  \institution{Beijing University of Posts and Telecommunications}
}

\author{Wen-Huang Cheng}
\affiliation{%
  \institution{National Taiwan University}
}

\author{Bei Liu}
\affiliation{%
  \institution{Microsoft Research Asia}
}

\author{Zhaoyang Zeng}
\affiliation{%
  \institution{International Digital Economy Academy}
}

\author{Jia Wang}
\affiliation{%
  \institution{National Yang Ming Chiao Tung University}
}

\author{Qiushi Huang}
\affiliation{%
  \institution{University of Surrey}
}

\author{Jiebo Luo}
\affiliation{%
  \institution{University of Rochester}
}

%

%

\renewcommand{\shortauthors}{Bo Wu et al.}

%
\begin{abstract}
Social Media Popularity Prediction (SMPP) is a crucial task that involves automatically predicting future popularity values of online posts, leveraging vast amounts of multimodal data available on social media platforms. Studying and investigating social media popularity becomes central to various online applications and requires novel methods of comprehensive analysis, multimodal comprehension, and accurate prediction. 

SMP Challenge is an annual research activity that has spurred academic exploration in this area. This paper summarizes the challenging task, data, and research progress.  
As a critical resource for evaluating and benchmarking predictive models, we have released a large-scale SMPD benchmark encompassing approximately half a million posts authored by around 70K users. The research progress analysis provides an overall analysis of the solutions and trends in recent years. The SMP Challenge website (www.smp-challenge.com) provides the latest information and news.
\end{abstract}


%
%

\ccsdesc[500]{Information systems~Multimedia and multimodal retrieval}
\ccsdesc[300]{Information systems~Online advertising}
\ccsdesc[500]{Human-centered computing~Social media}

%
\keywords{Social Multimedia, Visual Prediction, Popularity Prediction}

%

%
\maketitle

\section{Introduction}
\begin{figure}[t]
	\begin{center} 
		\includegraphics[width=1.0\linewidth]{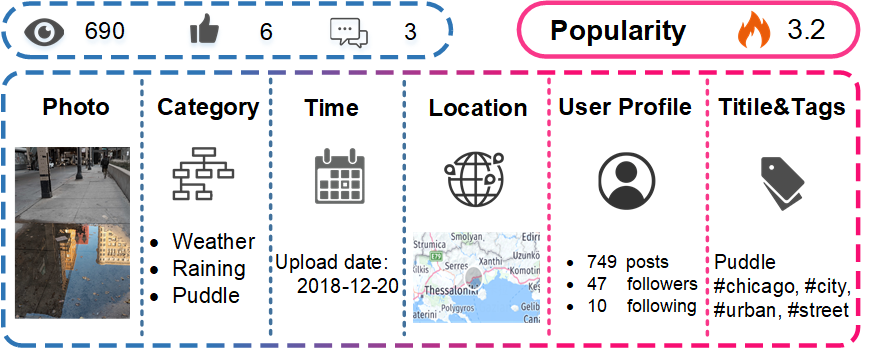} 
	\end{center}   
        \vspace{-0.15in} 
	\caption{Task: Social Media Popularity Prediction~\cite{Wu2019smp}. }
	\label{fig:post}
\end{figure}

On the ubiquity of social media (e.g., Facebook, Twitter, Flickr, YouTube, etc.), we are able to efficiently explore and share trending news, interesting photos or videos, fresh ideas, and valuable products or services. 
Social media platforms provide a crucial pathway for understanding and forecasting user preferences or behaviors in the real world.
Effectively learning the word-of-mouth of online posts from information oceans is the significant basis for further applications. 
Notably, Social Media Popularity Prediction (SMPP) is becoming emerging since it focuses on the forecasting task which facilitates various modern digital applications including demand forecasting, personalized advertising, social marketing, and social recommendation, etc. It benefits to enhance user experiences and improve business strategies for the current and future life.

SMP Challenge (Social Media Prediction Challenge) is an annual research challenge that aims to seek novel methods for forecasting problems and improving people’s social lives and business scenarios with numerous social multimedia data~\cite{Wu2017DTCN,Wu2016TemporalPrediction,Wu2019smp}. Meanwhile, the research interests are increased in studying rich social facts and knowledge with multi-modal information (e.g. images, text, video, events, etc.), while social media is now globally ubiquitous and prevalent. 
Since we hosted the activity for over 5 years, it attracted lots of participants teams, and research explorations.
The challenge built up the SMPD (Social Media Prediction Dataset), a large-scale, multimodal social media dataset including about 70K users, 500K online posts with temporal popularity covered over 500 days, user profiles, vision-language metadata, and 756 category tags. In addition, the dataset captured the social media popularity of each online post and records the post timestamps.

Social Media Popularity Prediction aims at predicting the future impact of an online post on a social media~\cite{Wu2017DTCN, Szabo2010predicting}. Existing work explored the social media popularity prediction as a time-related prediction problem~\cite{Shulman2016,Kobayashi2016}, and formulates popularity by online user attention or interaction (e.g. clicks, visits, reviews). 
To achieve this goal, the participating teams need to design new algorithms of understanding and learning techniques, and automatically predict by considering post content, future post time, and its multiple multimedia information (as shown in Figure~\ref{fig:post}) in a time-related dynamic system~\cite{Myers2014bursty,Kong2014predicting,Yang2011}.  
So far, the research on social media prediction covered several significant areas of multimedia and artificial intelligence and is closely integrated with computer vision, machine learning, natural language, and human-centered interaction.


\begin{figure}[t] 
	\begin{center}
 		\includegraphics[width=0.9\linewidth]{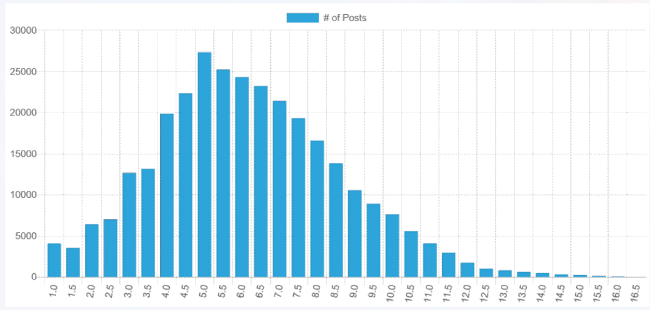} 
	\end{center}
        \vspace{-0.15in} 
	\caption{The overall distribution of popularity score over posts in the dataset SMPD. } 
	\label{fig:his}
\end{figure}
\section{Social Media Popularity Prediction}
Social Media Popularity Prediction (SMPP), introduced in early-stage works~\cite{Wu2017DTCN, Szabo2010predicting}, involves analyzing and learning from social media data. In this task, the popularity of online posts, such as photos, videos, or news articles, is influenced by the temporal characteristics of the social media system. These characteristics exhibit complex contexts and patterns, making accurate prediction of temporal popularity more challenging. Considering the changing popularity trends, the objective is to forecast the future impact of social media posts at a particular time before they are shared on the online platform.
Our task involves predicting the social media popularity $s$ of a new post $v$ from a user $u$ if it were to be published at a specific time $t$. The popularity score quantifies the level of interactions (such as views, likes, clicks, etc.) the post receives on a social multimedia platform. In our challenge, we consider the ``viewing count'' as a fundamental measure of post popularity, although this definition can be more comprehensive.

\textbf{Popularity Normalization.} To suppress the large variations among different photos (e.g. view count of different photos varies from zero to millions), we implement a log function~\cite{Wu2016TemporalPrediction} to normalize the value of popularity. The distribution is shown in Figure~\ref{fig:his}. 
The sequence of posts of a user, along with their corresponding timestamps, can be regarded as time-series data. The objective of the SMP Challenge is to create time-series feeds for predicting popularity. To achieve this, we defined the sequence data with a specific order of timestamps:

\textbf{User-Post Sequence~\cite{Wu2019smp}.} Suppose we have $n$ user-photo pairs and the sharing time of each pair. The user-post sequence can be denoted by $S = \{(u_1, v_1),(u_2, v_2), ..., $ $(u_n, v_n)\}$ with its sharing time order $t_1 \leq t_2 \leq ... \leq t_n$.

\section{Social Media Prediction Dataset}

\begin{figure}[t]
	\begin{center}
		\includegraphics[width=0.95\linewidth]{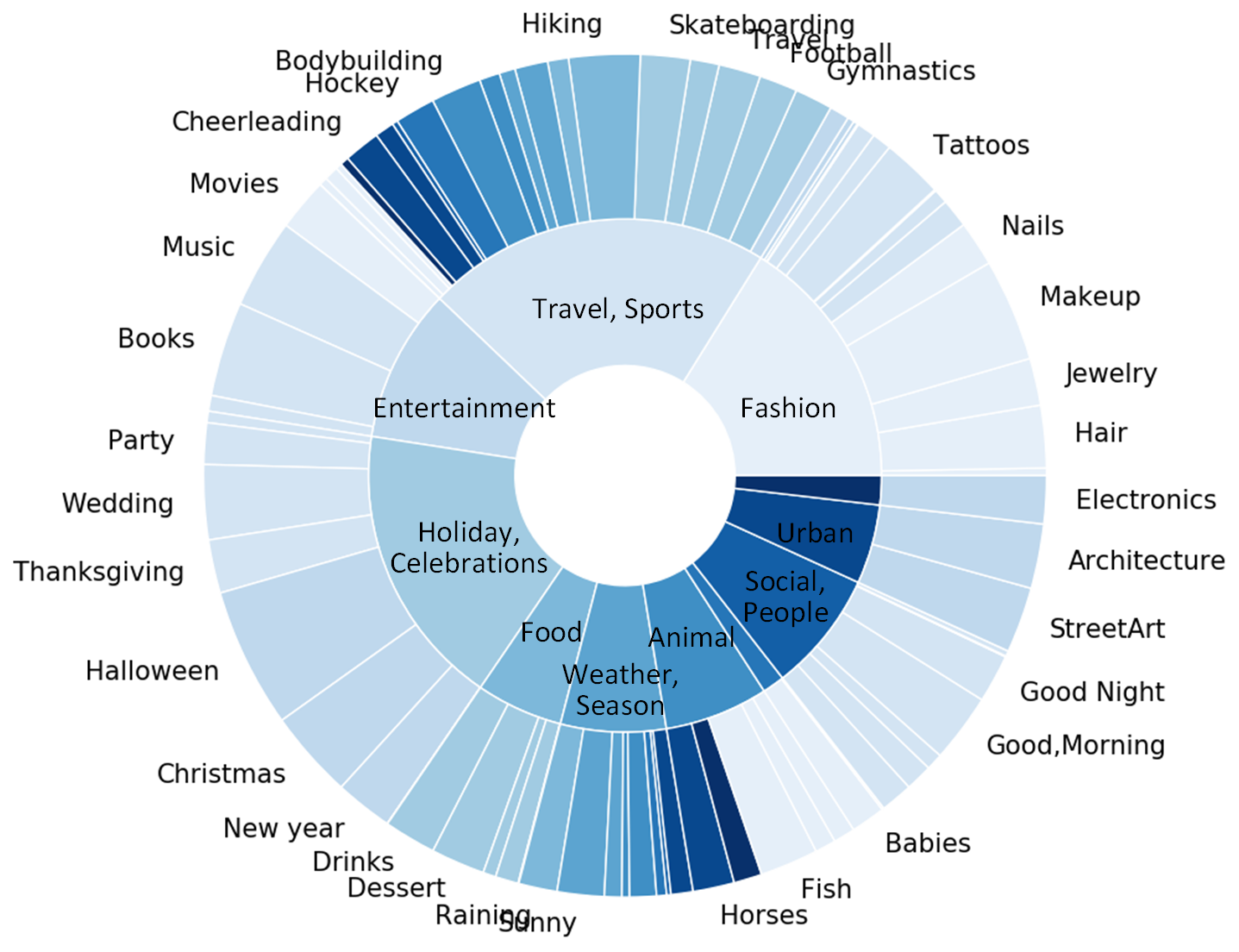} 
	\end{center}
	\caption{The statistics of $1^{st}$ and part of $2^{nd}$ level categories of photos. The width of each sector represents the percentages in respect of the total. }
	\label{fig:cate}
\end{figure}

Social Media Prediction Dataset (SMPD)~\footnote{http://smp-challenge.com/dataset} is a large-scale benchmark for social multimedia research. We collected the online posts and the multimodal information from Flickr which is one of the largest photo-sharing social media with over two billion photos monthly~\cite{Franck19}. Unlike single-task datasets, SMPD is a multifaceted collection that encompasses diverse contextual information and annotations, supporting multiple research tasks. The overview statistics of the dataset are shown in Table~\ref{tab:sta}. It contains over 486K posts from 69K online users. And each social media post has corresponding visual content and textual content information (e.g. posted photos, photo categories, custom tags, temporal and geography information). To create a multi-faced dataset for social media research, we meticulously categorized the post tags into 756 categories across 11 topics, as shown in Figure~\ref {fig:cate}. The data was collected daily from November 2015 to March 2016, covering a wide range of content types: visual content (e.g. photo and photo categories), textual content (e.g. post title and custom tags), and spatio-temporal content. These settings allow for the exploration of the impact of regions and time zones on online social behavior. In conclusion, SMPD represents a large-scale and multifaceted benchmark that enables extensive research in the field. 


\begin{table}
\centering
\caption{SMPD: Statistical Summary of the Dataset.} \label{tab:sta}
\begin{tabular}{l|c|c}%
\hline  %
\multirow{2}*{\textbf{Metrics}}&\multicolumn{2}{c}{ \textbf{Statistics}}\\
\cline{2-3}
&Train&Test\\
\hline  %
Number of Posts& $3.05\times10^5$&$1.81\times10^5$ \\
Average Popularity of Posts& 6.41&5.12\\
Popularity STD of Posts& 2.47&2.41\\
Number of users &$3.8\times10^4$&$3.1\times10^4$\\
\hline
Temporal Coverage of Posts& \multicolumn{2}{c}{480 days}\\
Number of custom tags& \multicolumn{2}{c}{$2.5\times10^5$}\\
Number of $1^{st}$ level categories& \multicolumn{2}{c}{11}\\
Number of $2^{nd}$ level categories& \multicolumn{2}{c}{77}\\
Number of $3^{rd}$ level categories&\multicolumn{2}{c} {668}\\
\hline %
\end{tabular}
\end{table}


\subsection{Visual Content}
Users find it easier to reflect their thoughts or emotions through photos/images on social media~\cite{Cappallo2015}. As introduced, we have 486k posts by 756 selected category keywords via APIs~\cite{FlickrAPI}. These keywords can be organized into 11 topics ranging from nature, people to animals (the directory tree in the left of Figure~\ref{fig:cate}).
Posts sharing the same keyword exhibit visual similarities in their content. 
Additionally, we have meticulously generated 668 individual $3^{rd}$ level human-crafted categories for the photos of posts. These hierarchical fine-grained annotations provide detailed and specific categorization, resulting in fine-grained classes of the posts.



\subsection{Spatio-temporal Content}

\subsubsection{Time}
Popularity prediction of social media posts is a time-sensitive task~\cite{Ellering2016}. The temporal context of posts records user activities, and it is necessary to identify the uploading time. Meanwhile, Flickr provides an uploading time for each submitted post.
The posts show that most of the posts in SMPD were uploaded between March 2015 to the creation time of the dataset in 2016. Among the upload dates, October and December become the popular month for sharing posts on the Flickr social media platforms. We attribute those improvements to the holidays.

\subsubsection{Location}
Location information provides user spatial distribution~\cite{Yang2016}. 
Although not all posts contain location data, the presence of location information in photos highlights the spatial regions of user activities. In SMPD, we have gathered about 32K posts with Point of Interest (POI) location information.
SMPD offers geo levels ranging from 1 to 16, which indicate the level of geo-locational granularity. These values span from world-level accuracy (level 1) to street-level accuracy (level 16), allowing for a fine-grained understanding of the location precision associated with each post.


\subsection{Textual Content}
In addition to visual content, we also collected the surrounding text of posts provided to show semantic information for each post. As statistics, there are more than 95\% of posts have relative descriptions or titles. When uploading a photo on the social media platform, the relative textual content is appended to provide more details about the photo content or publisher status. 

\textbf{Post Title.} Each posted photo has a unique title named by the user. As the saying goes: ``There are a thousand Hamlets in a thousand people$'$s eyes", each title contains an explanation and understanding of the photo. As shown in Table~\ref{tab:sta}, users utilize average 29 words to describe the content of uploaded photos, which not only helps to analyze the visual content of the posts but also relate to the popularity of the corresponding post.
\begin{figure}[t]
	\begin{center}
		\includegraphics[width=0.9\linewidth]{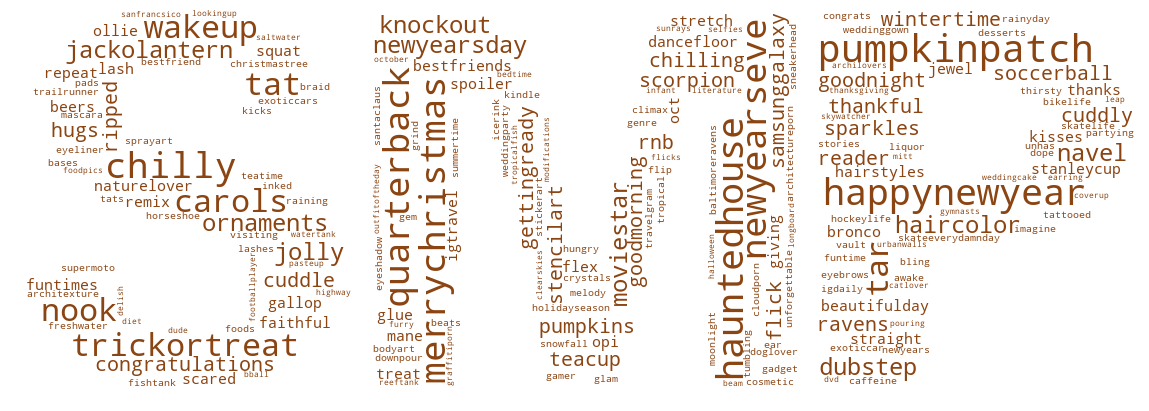} 
	\end{center}
	\caption{The tag-cloud of posts. The font size represents the tag frequency~\cite{Wu2019smp}.}
	\label{fig:cloud}
\end{figure}

\textbf{Post Tags.} Hashtags are used to help users easily find relevant content on specific interests or topics tagged with the same words. The advantage of this approach is that a single post can be labeled with multiple tags, making it more flexible than traditional topic channel separation. In Figure~\ref{fig:cloud}, we generated the ``tag cloud" of collected posts. By counting the tag frequently used, we can have a glimpse of which topics are more popular. The font size indicates the frequency. It shows that users prefer to share holiday posts. 

\section{Evaluation}

The combined use of SRC and MAE provides quantitative and objective evaluations of the performance of the models:
\begin{itemize}
\item Ranking Relevance: to measure the ordinal association between ranked predicted popularity scores and actual ones.
\item Prediction Error: to judge the error of the score prediction.
\end{itemize}

For quantitative performance evaluation, we will calculate two metrics: Spearman Ranking Correlation (SRC or Spearman's Rho) and Mean Absolute Error (MAE) for each submitted model. The SRC can be expressed as follows:
\begin{equation}
SRC= \frac{1}{k-1} \sum ^k _{i=1} \left( \frac{P_i - \bar{P}}{\sigma_P} \right) \left( \frac{{\hat{P}}_i - \bar{\hat{P}}}{\sigma_{\hat{P}}} \right), 
 \end{equation}
where $\bar{P}$ and $\sigma_P$ are mean and variance of the corresponding popularity set. In addition to SRC, we also employ Mean Absolute Error (MAE) to calculate the average prediction error, and it is determined as follows: 
\begin{equation}
MAE=\frac{1}{k}\sum_{i=1}^{n}\mid\hat{P}_{i}-P_{i}\mid.
 \end{equation}
 
The competition's ranking relies on an objective evaluation method. To determine the rankings, teams are assessed based on specific objective evaluation metrics, and their performance on each metric is used to create a rank. The final rank of a team is determined by combining its rankings on the individual metrics in a balanced manner. 

\section{Methods for the SMP Challenge}
Drawing from the methodologies implemented by past award-winning teams in the SMP challenge, we have compiled a comprehensive overview of the methodologies employed in the competition up to the present year.
These methodologies can be broadly classified into four categories, including Ensemble leaning-based, Convolutional Neural Network (CNN)-based, Recurrent Neural Network (RNN)-based, and Attention-based approaches.

\subsection{Ensemble Learning}
It is common to utilize ensemble learning algorithms, which involve combining multiple weak classifiers or sub-classifiers to create a stronger classifier. These approaches, such as bagging, boosting, and bootstrap strategies (e.g., SVM~\cite{hearst1998svm}, LightGBM~\cite{ke2017lightgbm}, logistic regression~\cite{collins2002logistic}, classification trees~\cite{liaw2002classification}, and XGBoost~\cite{chen2016xgboost}), are widely used to improve predictive accuracy.
In the SMP competition, most advanced techniques utilize ensemble learning as the final regressor to handle texture, visual or fused features, enabling accurate prediction of post popularity. In earlier approaches, a single model was typically used for handling fused texture and visual features. Recently, however, separate texture and visual information ensemble models have become more prevalent, and weighted algorithms are employed to obtain the final prediction score.

\subsection{Convolutional Neural Networks}
Convolutional Neural Networks (CNNs) have emerged as a highly effective category of deep learning models renowned for their exceptional performance in computer vision applications. The key strength of CNNs lies in their ability to extract high-level features from raw input data, primarily accomplished through the utilization of convolution layers. This characteristic has contributed to their remarkable achievements in tasks such as image classification, object detection, and semantic segmentation.
In the SMP Challenge, a significant number of works~\cite{xu2020multimodal, hsu2020rethinking, hsu2019popularity, chen2019social} have adopted CNN models to extract visual information from the posts. Notably, popular CNN architectures like ResNet-101\cite{he2016deep} are frequently employed. This approach allows models to learn the visual feature representations of CNNs to effectively capture the multiple-scale visual characteristics present in the post images.

\subsection{Recurrent Neural Networks}
Recurrent Neural Networks (RNNs) have gained significant prominence as a category of artificial neural networks extensively employed in the modeling of sequences or time-series data. Their distinctive characteristic lies in their ability to handle and analyze sequential data effectively. This capability stems from their capacity to maintain a hidden state that captures and retains information from preceding inputs.
RNNs are capable of modeling intricate dependencies and temporal relationships inherent in sequential data. As a result, RNNs have proven to be highly valuable for tasks such as speech recognition and natural language processing, where understanding and processing sequential information are crucial. The inherent nature of RNNs allows them to excel in these domains, enabling the capture and analysis of the sequential patterns and dynamics present in the data.
Thus, in order to process the time-sensitive information in SMP, some teams leverage RNN models~\cite{chen2020curriculum, hsu2019popularity, hsu2020rethinking} to capture the relevant features from continuous date data. By incorporating RNN models into SMP techniques, teams can better predict the future popularity of posts based on their historical performance.

\subsection{Attentions and Transformers}
Over the course of 2019 to 2023, the SMP challenge has seen active participation from researchers who have continuously integrated attention algorithms and transformer modules into their methods.

In 2019, several works utilized attention-based models. For instance, ~\cite{ding2019social} applied a pre-trained BERT model to extract 768 dimensional text features. This approach surpassed conventional methods like one-hot encoding and shallow Word2Vec models~\cite{mikolov2013efficient} by enabling the BERT model to grasp the contextual information within each post.
After that, in 2020, several early-stage attempts incorporated attention strategies in their methods. One of them, \cite{chen2020curriculum}, combined curriculum learning with transformer models. They trained their model on progressively complex samples and leveraged the trained model to extract texture and image features. Additionally, the RoBERTa model~\cite{liu2019roberta} was employed to extract textual embedding.
Then, in 2022, most of the participants leverage transformer modules to extract high-dimensional features. Among them, \cite{chen2022title} adopted a combination of visual and language transformer models. Furthermore, they devised a title-label contrastive learning method to obtain both title-visual and label-visual information. 
In the context of ongoing yearly advancements, participants have been introducing interesting and innovative strategies that seamlessly integrate attention mechanisms into social media prediction tasks. These new approaches encompass a variety of techniques, such as aligning visual and textual information, and the generation of additional paired learning training data. These innovative methods have demonstrated promising potential for enhancing the accuracy and effectiveness of social media prediction models.

As shown in Table\ref{tab:baseline}, the approach outlined in the previous summary demonstrates a progressive evolution, starting with the use of pre-trained models to extract text features, followed by a transition to models capable of extracting both text and image features. Subsequently, fine-tuning the model with data and creating specialized alignment or contrastive learning models were incorporated.
These methods have yielded significant improvements in performance, as evidenced by the leaderboard on the official website. Notably, the top-1 team's SRC (Success Rate of Correct Prediction) increased substantially from 0.59 to an impressive 0.77. The results underscore the effectiveness of the attention model in feature extraction for both text and image data, leading to enhanced predictive capabilities.

\begin{table}
\centering
\caption{Baseline Approaches on SMPD.} \label{tab:baseline}
\begin{tabular}{l|c|c|c}%
\hline  %
\textbf{Method}&\textbf{Regression}&\textbf{SRC}&\textbf{MAE}\\
\hline %
ResNext-101~\cite{xie2017aggregated} & LightGBM & 0.322 & 1.87 \\
BERT(Base)~\cite{devlin2018bert} & LightGBM & 0.538 & 1.65 \\
GPT-2~\cite{radford2019language} & LightGBM & 0.574 & 1.59 \\
FastText~\cite{bojanowski2017enriching} & CatBoost & 0.622 & 1.49 \\
FastText~\cite{bojanowski2017enriching} + ResNext~\cite{xie2017aggregated}   & CatBoost& 0.70 & 1.40 \\
\hline %
Winner2020-A~\cite{wang2020feature} & CatBoost & 0.674 & 1.36 \\
Winner2020-B~\cite{lai2020hyfea} & CatBoost & 0.704 & 1.37 \\
Winner2022~\cite{wu2022deeply} & CatBoost & 0.71 & 1.24 \\
\hline %
\end{tabular}
\end{table}
\section{Conclusions}

In this paper, we have presented an overview of the SMP Challenge and proposed a large-scale social multimedia dataset for real-world prediction challenges. Meanwhile, we formulate the temporal popularity prediction task, analyze the proposed dataset and define evaluation metrics.

%

%
\newpage
\bibliographystyle{ACM-Reference-Format}
\balance
\bibliography{main} 

%

\end{document}